\documentclass[aps,prl,twocolumn,nofootinbib]{revtex4-1}
\usepackage[english]{babel}
\usepackage[utf8]{inputenc}
\usepackage[pdftex]{graphicx}
\usepackage{hyperref}
\usepackage{url}
\usepackage[titletoc,title]{appendix}
\usepackage[suffix=]{epstopdf}
\usepackage{amsmath}
\usepackage{amssymb}
\usepackage{dsfont}
\usepackage{mathtools}
\usepackage{bbm}

\newcommand{\erf}{\text{erf}}
\newcommand{\tr}{\text{tr}}

\newcommand{\figI}{
	\begin{figure*}[ht!]
		\centering
		\includegraphics[width=0.47\textwidth]{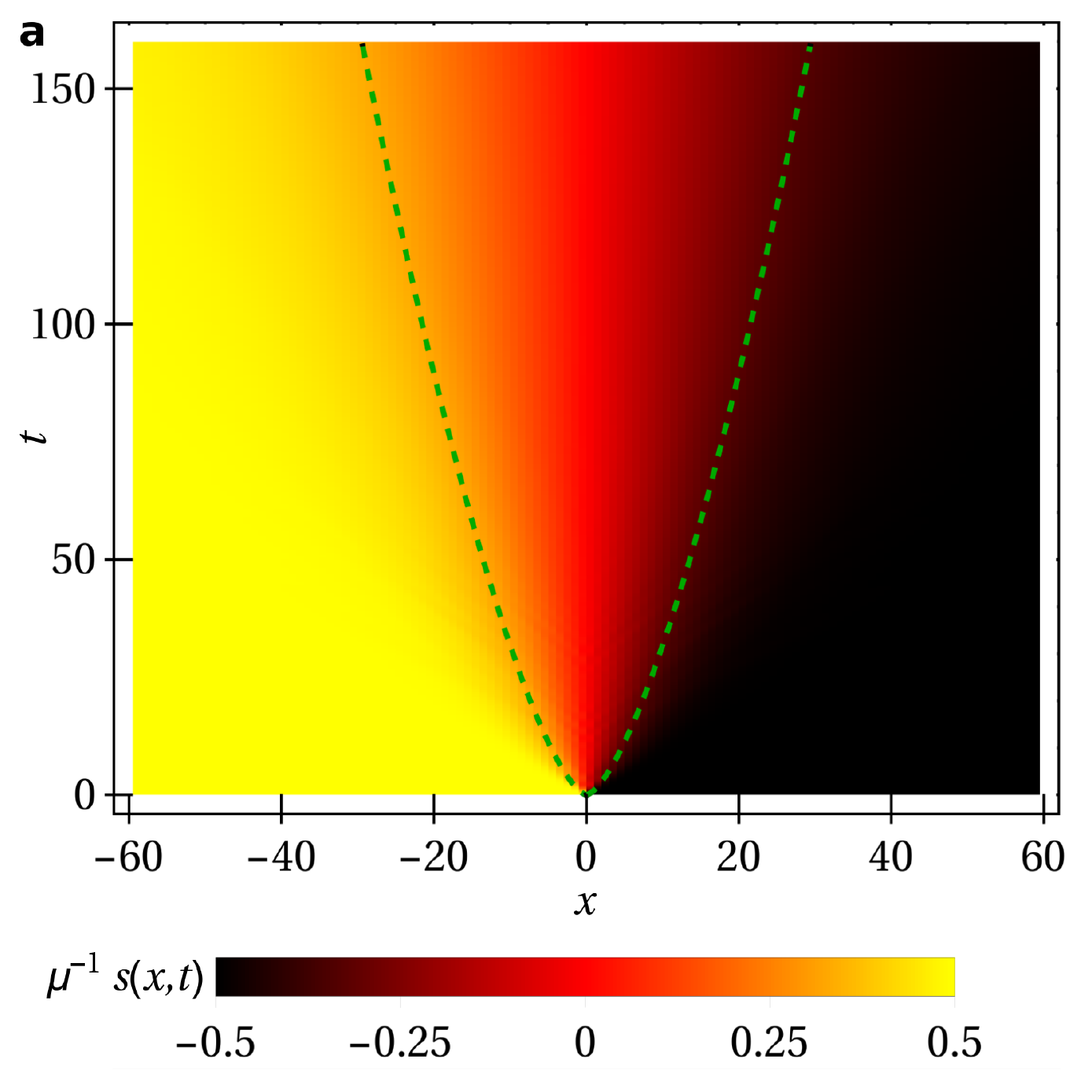}
		\includegraphics[width=0.47\textwidth]{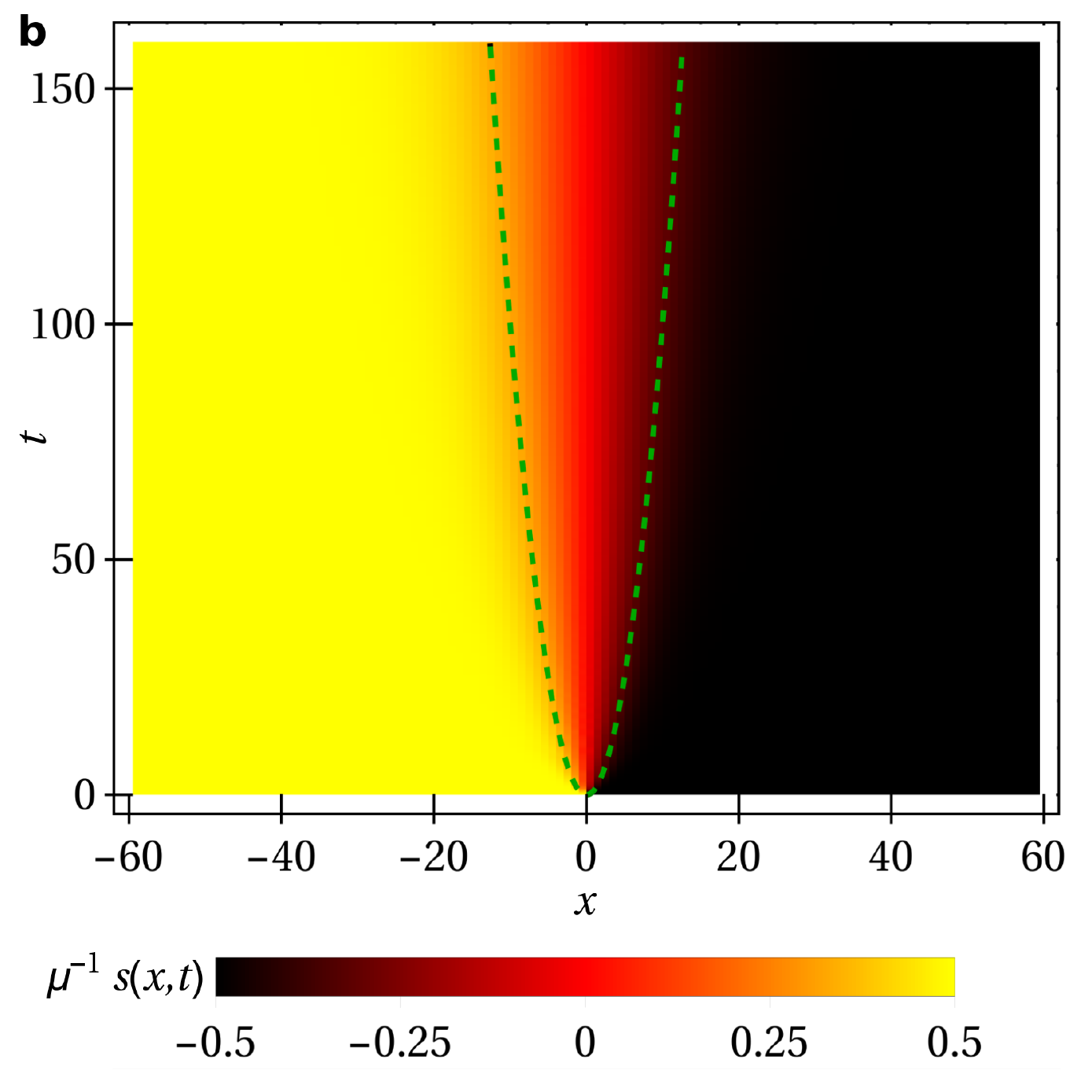}
		\includegraphics[width=0.47\textwidth]{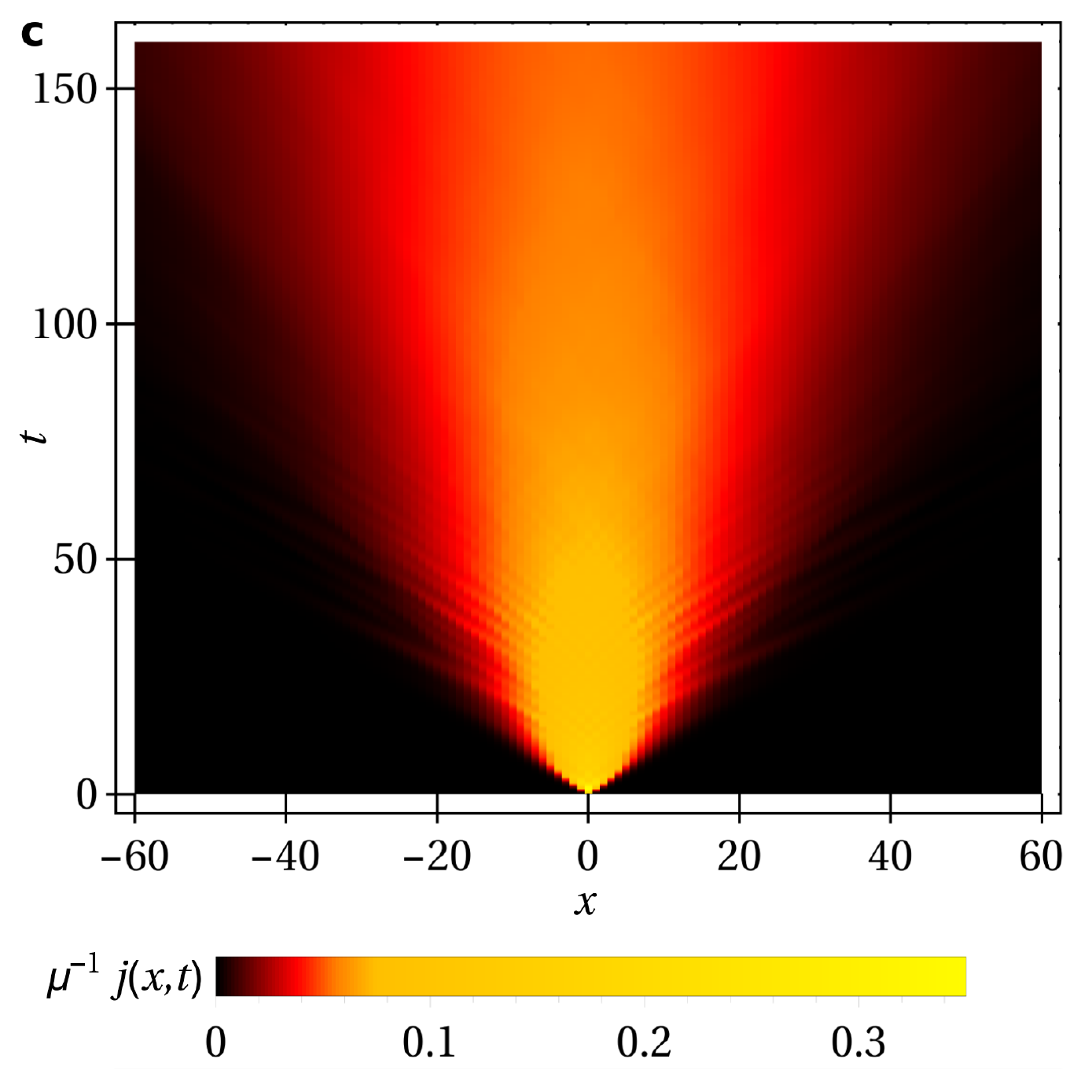}
		\includegraphics[width=0.47\textwidth]{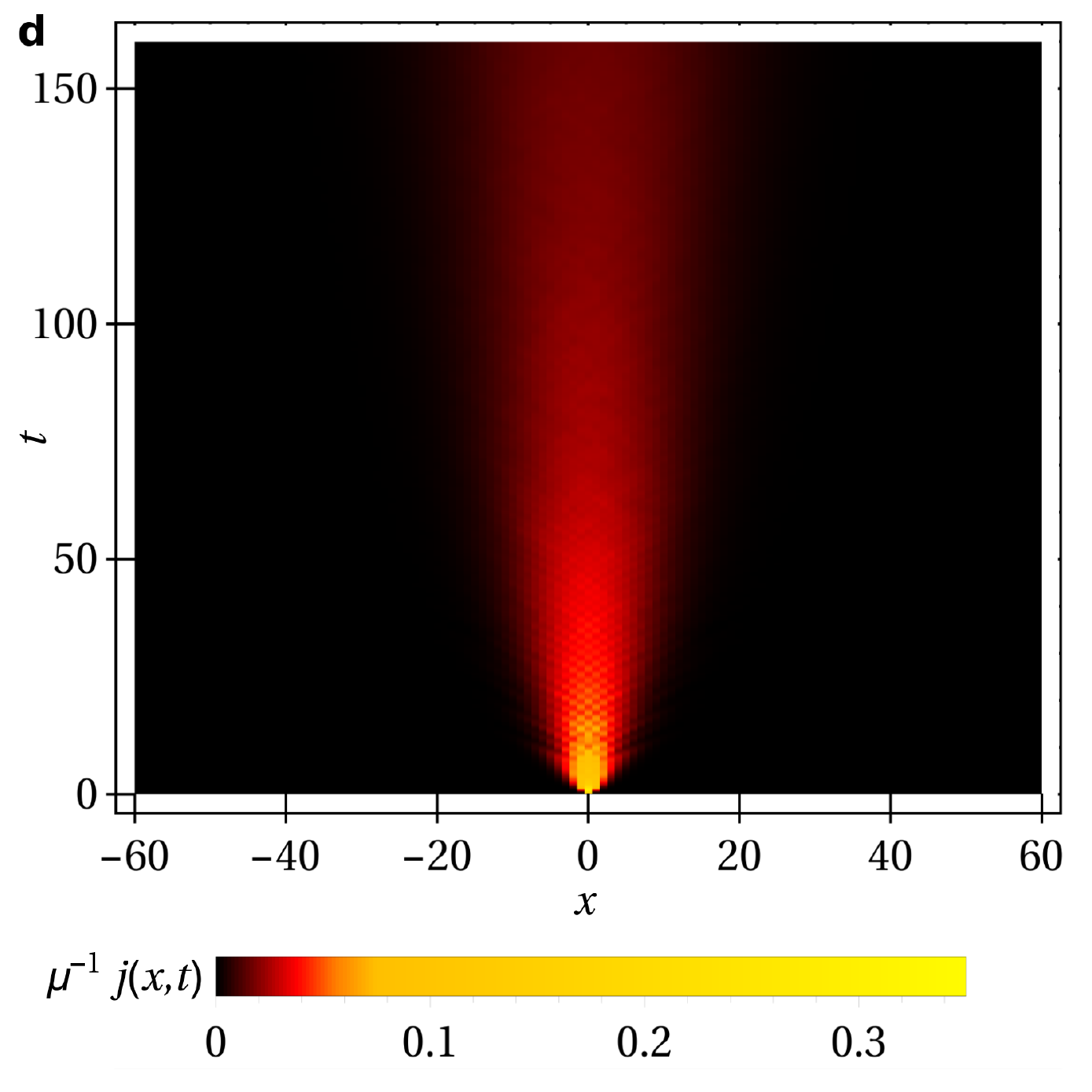}
		\caption{{\bf Dynamics of spin and current densities.} Time evolution of spin density $s(x,t)=\tr(\rho(t)s_x^{(3)})$ ((a) and (b)) and current ((c) and (d)) profile $j(x,t)=\tr(\rho(t)j_x)$ for the isotropic point $\Delta=1$ ((a) and (c)), and $\Delta=2$ ((b) and (d)), following an inhomogeneous quench. One can see that the spreading is much faster for $\Delta=1$, in both cases though it is slower than ballistic. Dashed green curves guide the eye towards scaling $x \sim t^{2/3}$ in (a), and $x \sim t^{1/2}$ in (b). Data are shown for $n=320$ and small initial polarisation $\mu=\pi/1800$.}
		\label{fig1}
	\end{figure*}
}

\newcommand{\figII}{
	\begin{figure*}[ht!]
		\centering
		\includegraphics[width=0.47\textwidth]{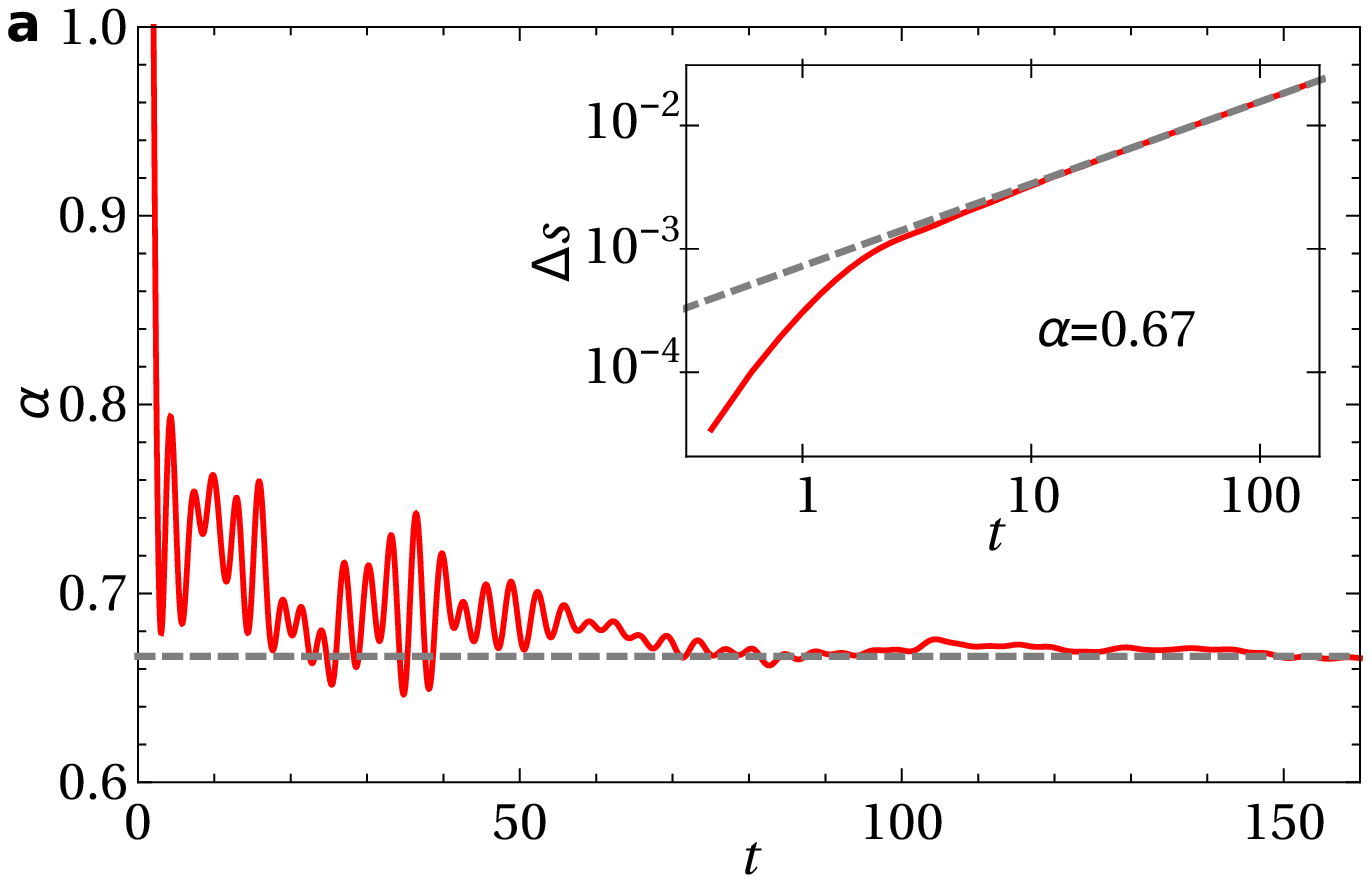}
		\includegraphics[width=0.47\textwidth]{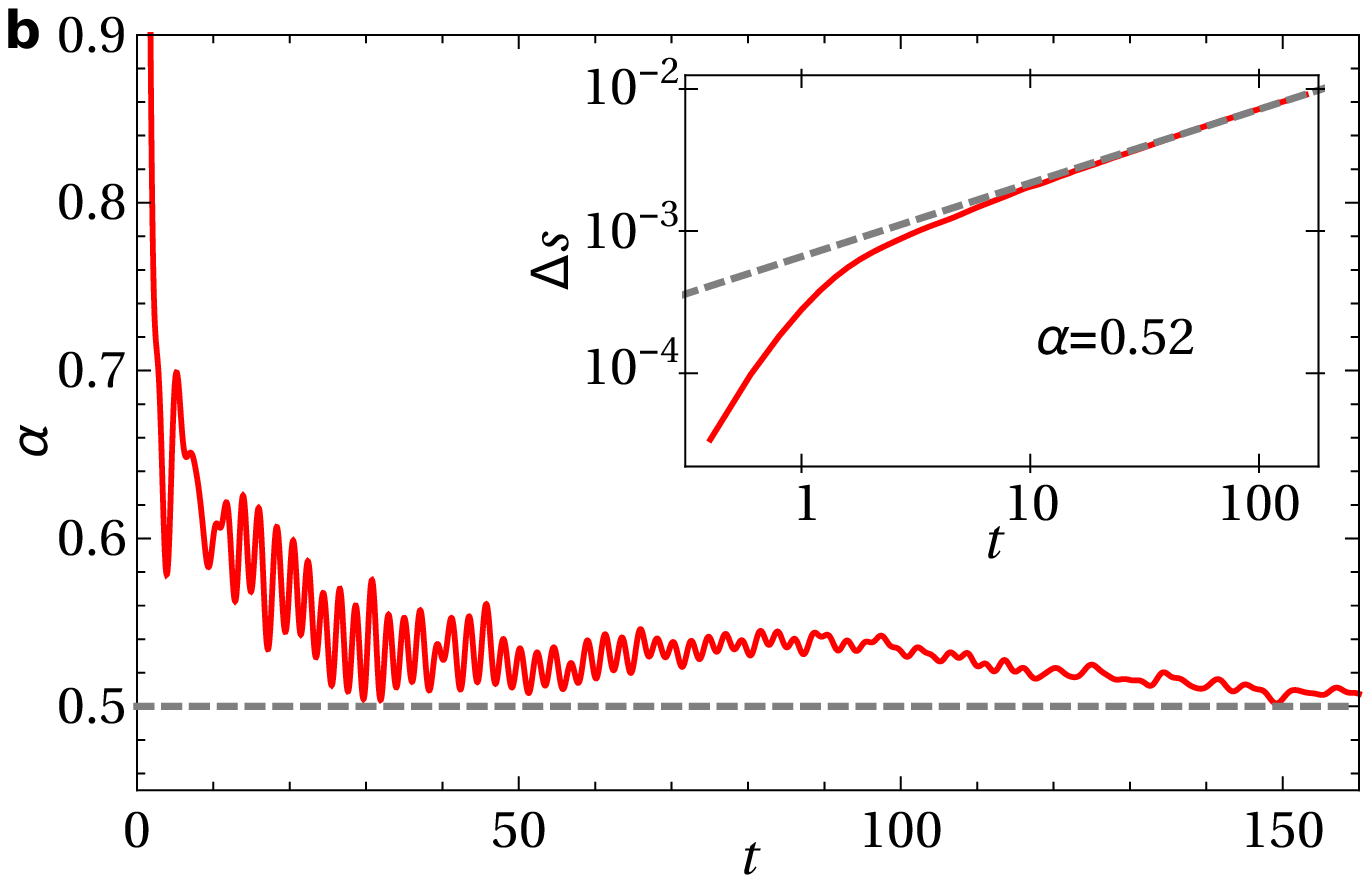}
		\includegraphics[width=0.47\textwidth]{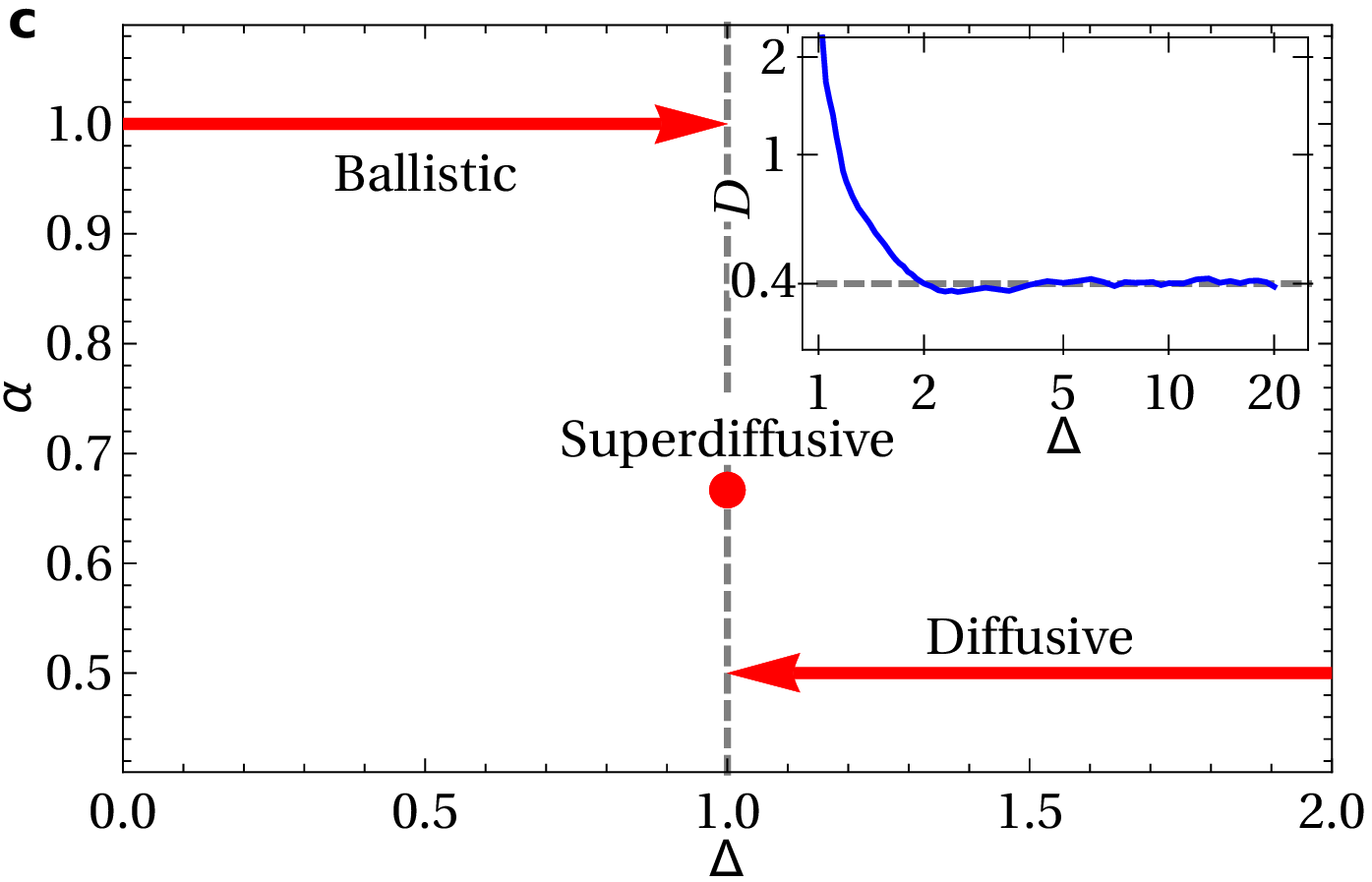}
		\includegraphics[width=0.47\textwidth]{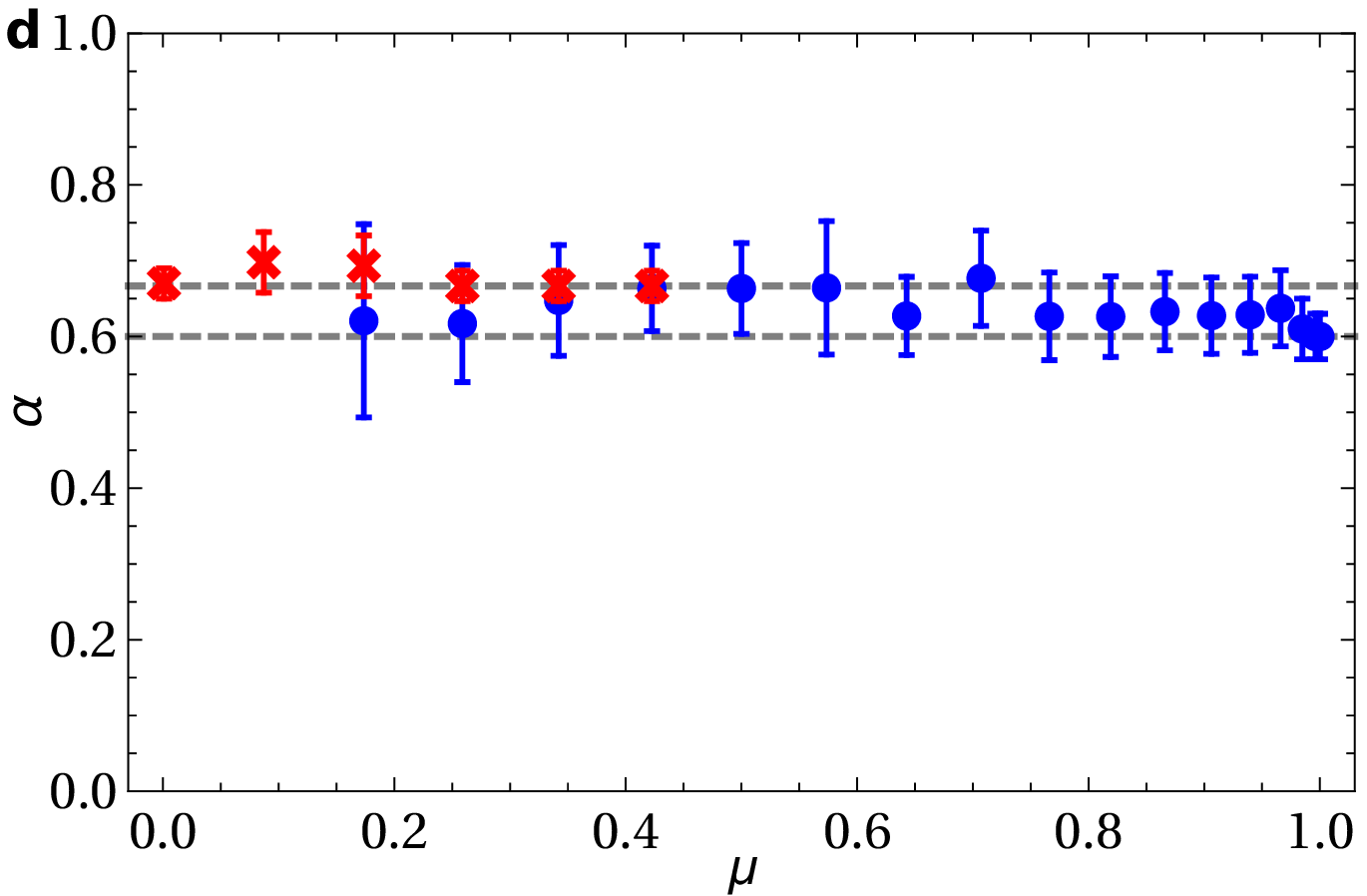}
		\caption{{\bf Scaling exponents of magnetisation spreading.} (a, b) Local exponent $\alpha(t)$ calculated as a numerical log-derivative ${\rm d}\log\Delta s(t)/{\rm d}\log t$ for $\Delta=1$ (a) and $\Delta=2$ (b) 
			(dashed lines indicate exponents $2/3$ and $1/2$, resp., while dashed lines in the insets show best power-law fits to $\Delta s(t)$ -- red curve), both for $\mu=\pi/1800$. (c) Conjecture for the dependence $\alpha(\Delta)$ at high temperatures and small $\mu$. The inset shows the diffusion constant obtained from Fick's law for various values of $\Delta$ in the diffusive regime, converging to a finite value at large $\Delta$~(agreeing with Ref.~\cite{FHM}). (d) Dependence on $\mu$ for $\Delta=1$ shows a small but significant change in the behaviour: for $\mu\approx 1$ it is closer to $\alpha=3/5$ while for small $\mu$ it becomes significantly close to $\alpha=2/3$ (dashed). The blue (circles) and red (crosses) symbols represent wave function and density operator evolutions respectively. We average over samples of $10-130$ random initial wave-functions for each blue data point. For intermediate $\mu$ the error-bars (denoting the estimated standard deviation) are larger since the simulation is less efficient in that regime (see Methods). }
		\label{fig2}
	\end{figure*}
}

\newcommand{\figIII}{
	\begin{figure*}[ht!]
		\centering
		\includegraphics[width=0.47\textwidth]{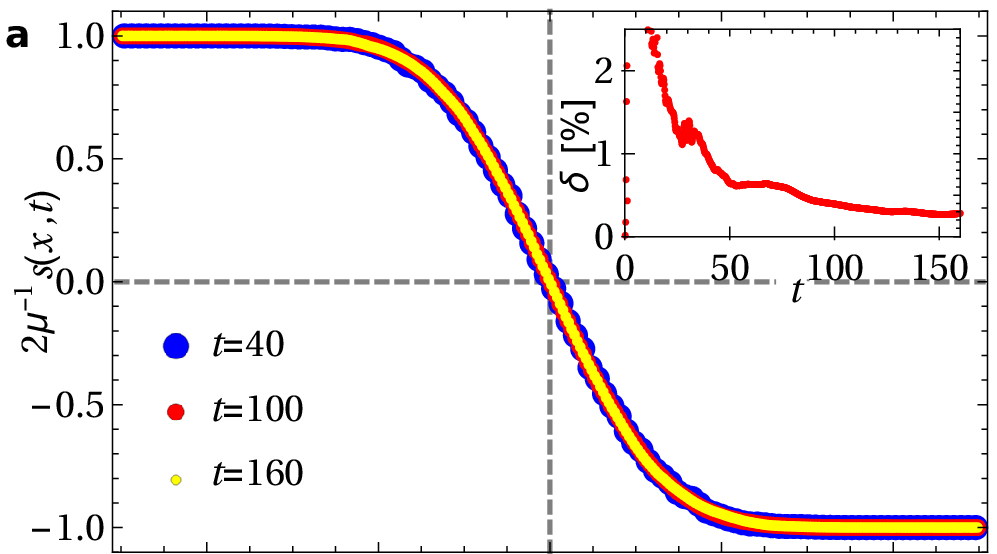}
		\includegraphics[width=0.47\textwidth]{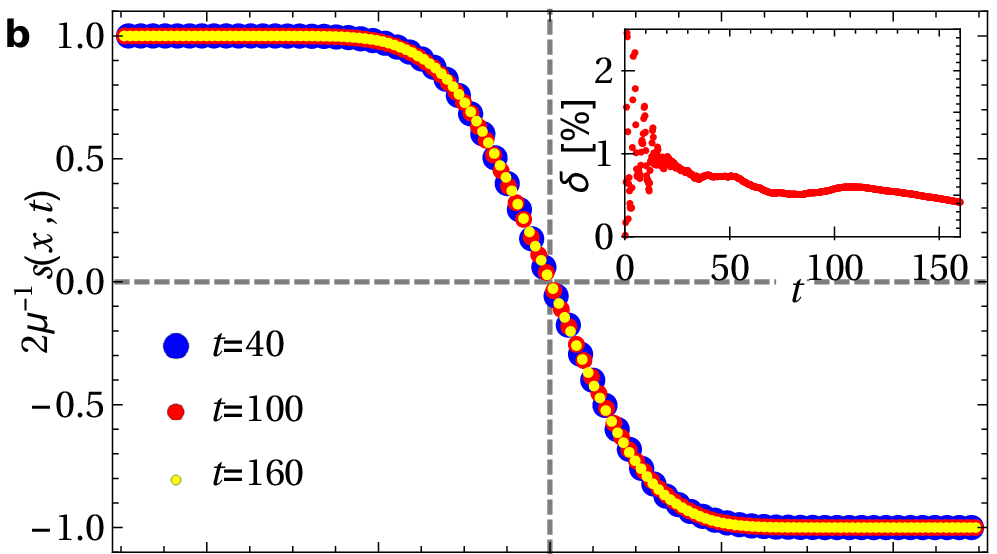}
		\includegraphics[width=0.47\textwidth]{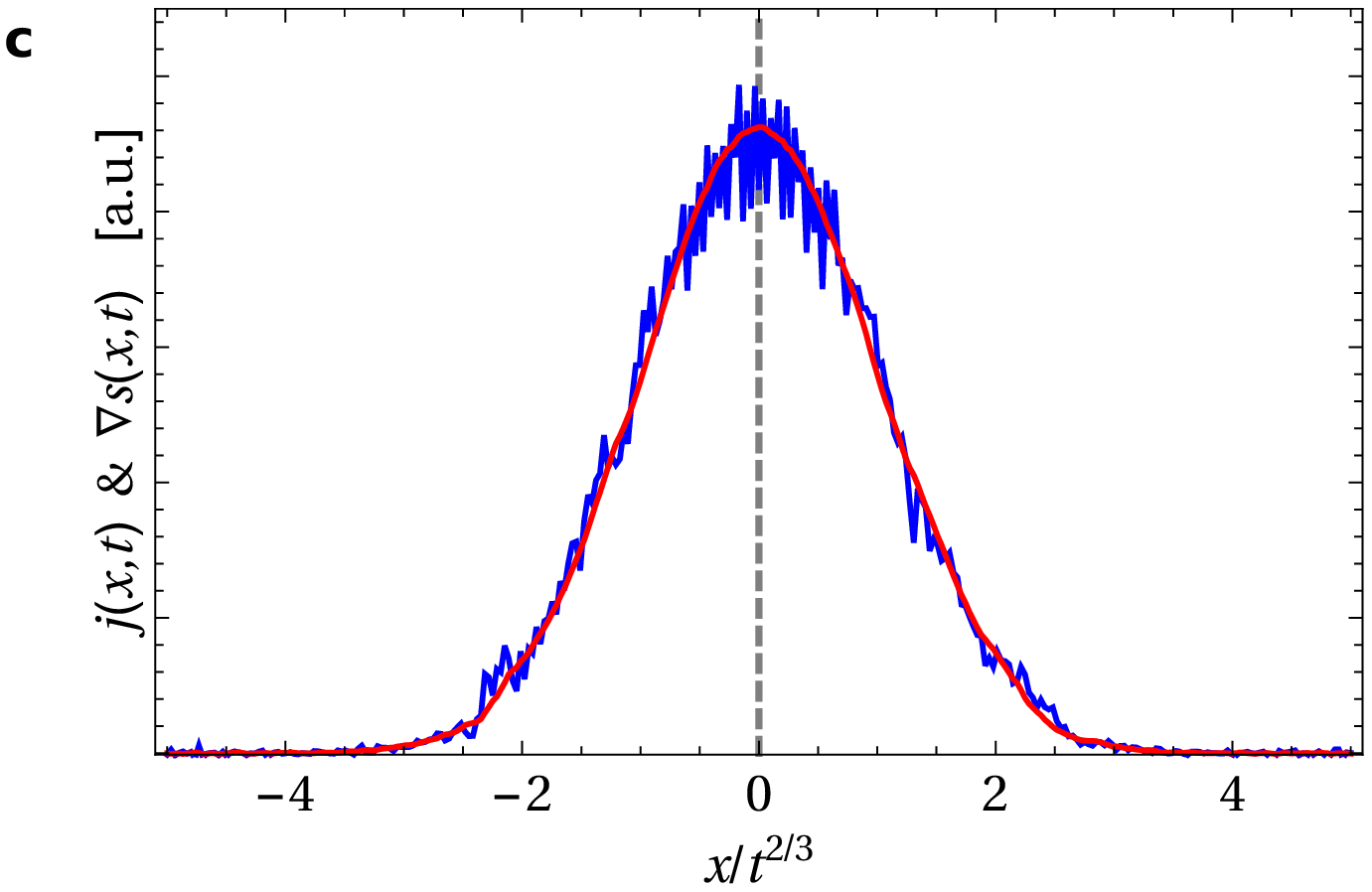}
		\includegraphics[width=0.47\textwidth]{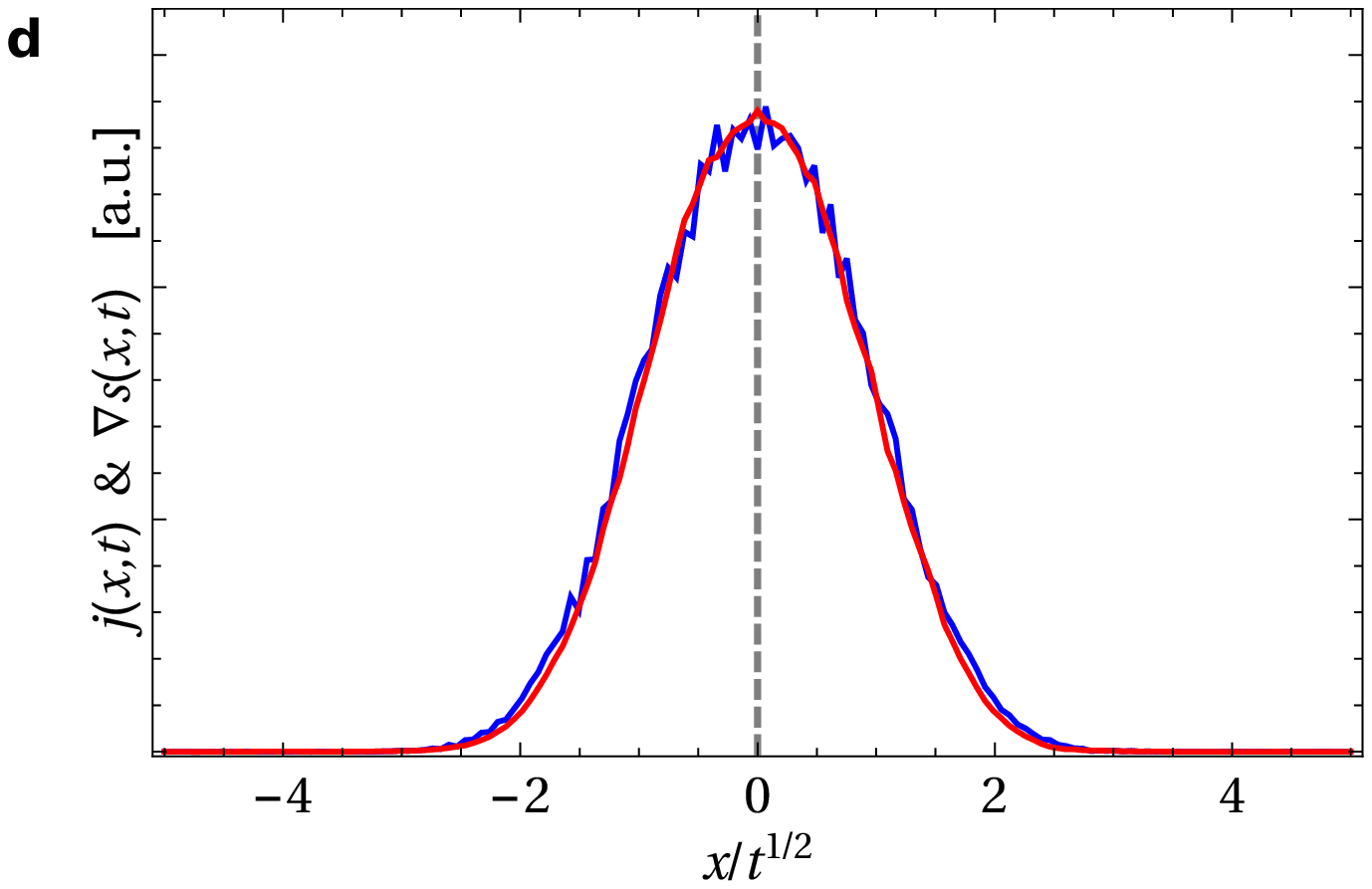}
		\caption{{\bf Scaling profiles.} Scaling of density and current profiles with $x/t^\alpha$. In panels (a) and (b) we show the scaling of magnetisation profiles, (a) for $\Delta=1$ using $\alpha=2/3$, and (b) for $\Delta=2$ and using $\alpha=1/2$ (note that the points for different times overlap almost perfectly; the insets show the convergence of the relative root-mean-square difference (in \%) between data $s(x,t)$ and scaled $\erf$-profiles (see text) as a function of time). 
			Frames (c) and (d) show the emergence of Fick's law at late times (shown at $t=160$), comparing current profiles (red) to gradients of spin density (blue) -- both indistinguishable from Gaussians,
			for $\Delta=1$ in (c) and $\Delta=2$ in (d). In all plots the system size is $n=320$.}
		\label{fig3}
	\end{figure*}
}

\newcommand{\figIV}{
	\begin{figure}[ht!]
		\centering
		\includegraphics[width=0.47\textwidth]{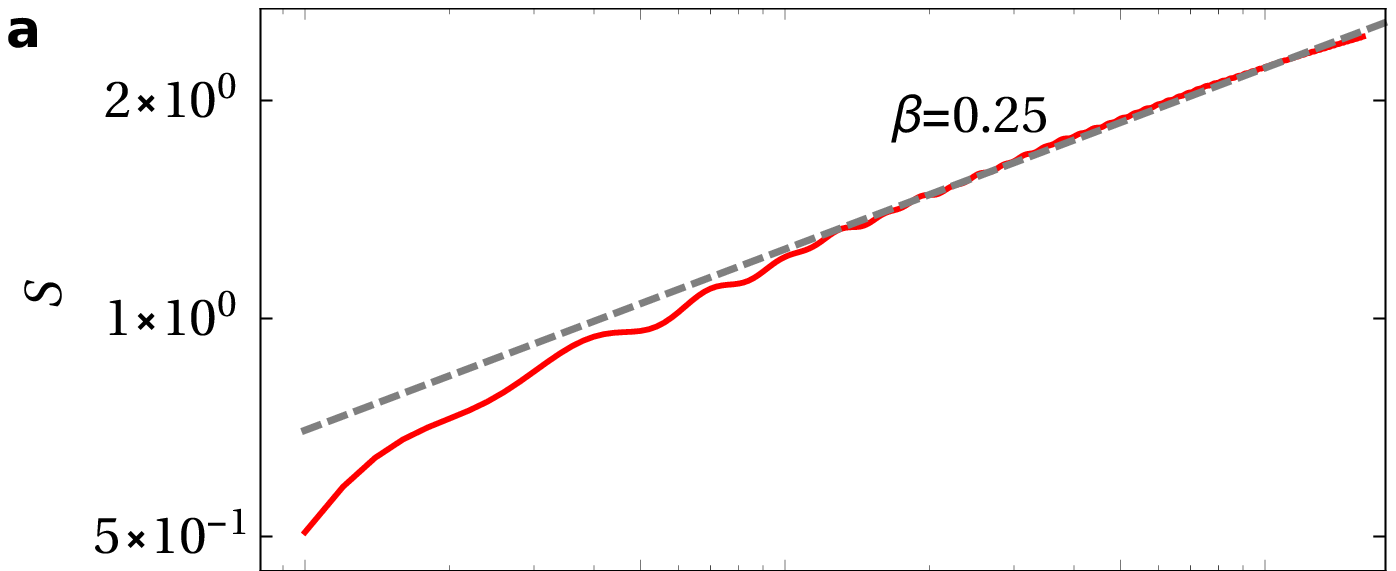}
		\includegraphics[width=0.47\textwidth]{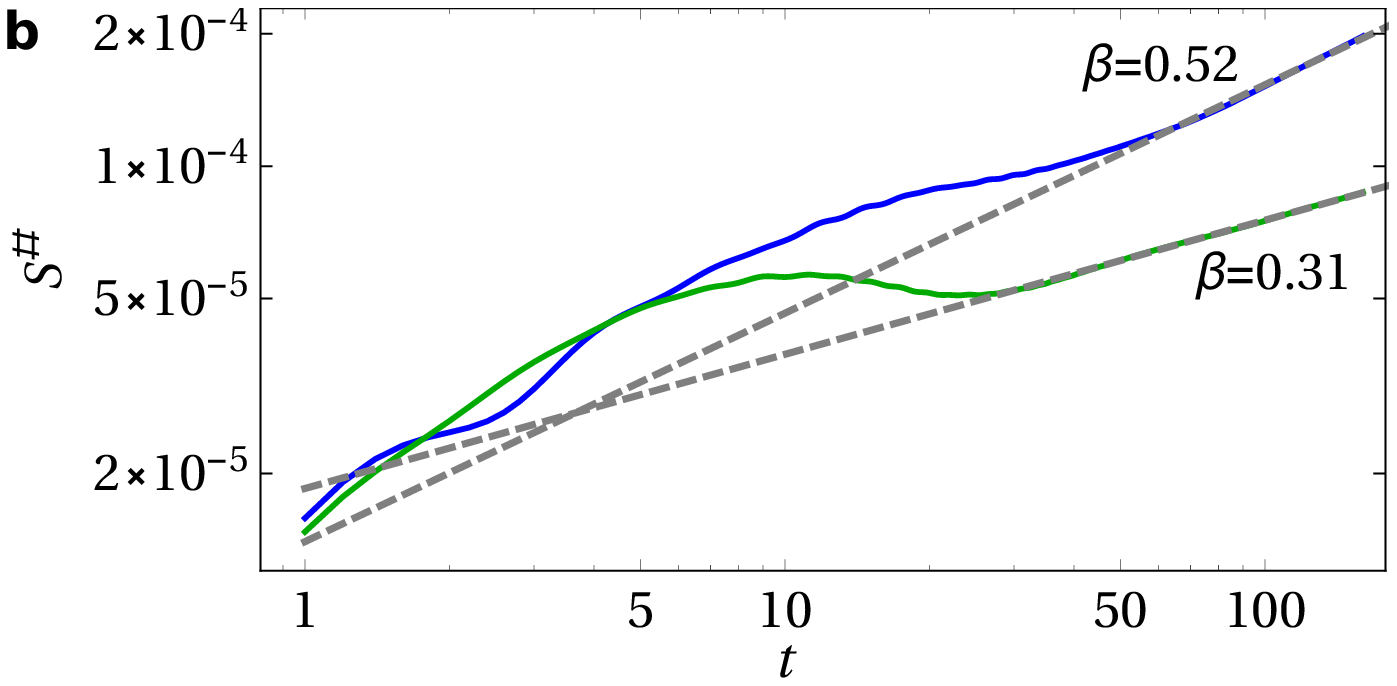}
		\caption{{\bf Simulation complexity.} (a) Von Neumann entanglement entropy $S$ for the fully polarized initial state ($\mu=1$) at the isotropic point $\Delta=1$. (b) Operator space entanglement entropy $S^{\#}$ for $\Delta=1$ (blue) and $\Delta=2$ (green), both for $\mu=\pi/1800$. Bipartition into two equal halves and a system size of $n=320$ are used.
		}
		\label{fig4}
	\end{figure}
}

\begin{document}
	
	\title{Spin diffusion from an inhomogeneous quench in an integrable system}
	
	\author{Marko Ljubotina}
	\affiliation{Physics Department, Faculty of Mathematics and Physics, University of Ljubljana, Jadranska 19, SI-1000 Ljubljana, Slovenia}
	\author{Marko \v Znidari\v c}
	\affiliation{Physics Department, Faculty of Mathematics and Physics, University of Ljubljana, Jadranska 19, SI-1000 Ljubljana, Slovenia}
	\author{Toma\v z Prosen}
	\email{tomaz.prosen@fmf.uni-lj.si}
	\affiliation{Physics Department, Faculty of Mathematics and Physics, University of Ljubljana, Jadranska 19, SI-1000 Ljubljana, Slovenia}
	
	\begin{abstract}
		Generalised hydrodynamics predicts universal ballistic transport in integrable lattice systems when prepared in generic inhomogeneous initial states. However, the ballistic contribution to transport can vanish in systems with additional discrete symmetries.	Here we perform large scale numerical simulations of spin dynamics in the anisotropic Heisenberg $XXZ$ spin $1/2$ chain starting from an inhomogeneous mixed initial state which is symmetric with respect to a combination of spin-reversal and spatial reflection. In the isotropic and easy-axis regimes we find non-ballistic spin transport which we analyse in detail in terms of scaling exponents of the transported magnetisation and scaling profiles of the spin density. While in the easy-axis regime we find accurate evidence of normal diffusion, the spin transport in the isotropic case is clearly super-diffusive, with the scaling exponent very close to $2/3$, but with universal scaling dynamics which obeys the diffusion equation in nonlinearly scaled time. 
	\end{abstract}
	
	\maketitle
	
	\section{Introduction}
	
	Integrable models, such as the classical Kepler problem, harmonic oscillators, the planar Ising problem, etc., form cornerstones of our understanding of nature. Their equilibrium physics is usually well understood, even for the most complicated among integrable models, e.g. the ones solvable by the Bethe ansatz~\cite{takahashi}. Nonequilibrium physics of quantum systems on the other hand is much less understood~\cite{RMP}, particularly when going beyond the simplest integrability of quadratic models. This theoretical gap is becoming even more apparent with the advancement of experimental methods that are offering us analog simulation of models beyond the capability of our best theoretical and numerical methods~\cite{Bloch,exper}. 
	
	Nonequilibrium dynamics of integrable quantum systems is thus one of the main current focuses of both theoretical and experimental condensed matter physics \cite{JSTAT}.	A macroscopic number of conservation laws existing in such systems \cite{qloc_review} provide a variety of ways to break ergodicity, manifesting, for instance, in equilibration processes to non-thermal states or ballistic high-temperature transport of conserved quantities, such as energy, magnetisation or charge. A naive classical reasoning might be that, because integrable systems are distinguished by constants of motion that force the dynamics to be simple and almost periodic (e.g., orbits winding up the torus), one should expect to see ballistic transport. We shall demonstrate that this picture, while being correct for trivially integrable noninteracting models, such as harmonic oscillator chains \cite{Lieb}, can in fact be wrong for an interacting quantum integrable model. 
	
	Recently, a generalisation of hydrodynamics has been put forward \cite{doyon,bertini} which successfully predicts ballistic currents and scaled density profiles of integrable interacting systems quenched from inhomogeneous initial states \cite{ruelle,bernard,bhaseen,enej,moore,doyonspohn}, which is a convenient method to study relaxation and nonequilibrium transport. In this protocol, the system is prepared in the state where the left and the right part, for $x<0$ and $x>0$ respectively, are in different equilibrium states, and then, at $t=0$, let to evolve with a homogeneous interacting Hamiltonian. However, when ballistic transport is prohibited due to generic symmetries, such as is the case for spin transport in the anisotropic Heisenberg spin chain in the easy-axis (Ising) regime, this theory makes no prediction. 
	
	In extended interacting integrable system a macroscopic number of local conservation laws exists, in number proportional to the number of degrees of freedom, which can be exploited to develop generalised hydrodynamics \cite{doyon}. This theory for typical inhomogeneous initial states predicts ballistic scaling $f(\xi=x/t)$ of densities and currents of conserved quantities, such as energy, charge or magnetisation. However, in systems with parity ($\mathbb{Z}_2$) symmetries, such as particle-hole exchange (or spin reversal), observables that are odd under the parity and initial states that are symmetric under the combined parity and spatial reflection $x\to-x$, the ballistic contribution to transport can vanish.  In fact, vanishing ballistic transport channel can then be related to the absence of local or quasi-local conserved charges with odd parity \cite{prosen,qloc_review}. This means that the transported conserved quantity at $x=0$ grows slower than linear with $t$.  
	
	Here we propose a conjecture, based on large scale simulations, that a quench from an inhomogeneous initial state will in such cases generically result in diffusive  spin dynamics. We demonstrate our results on the anisotropic Heisenberg chain ($XXZ$ model). However, we stress that the $XXZ$ model goes beyond being a mere toy model -- it has been instrumental in the development of quantum integrability \cite{Bethe,Baxter} and describes interaction in real spin chain materials \cite{Hlubek}. Remarkably, in the case of isotropic Heisenberg interaction, spin relaxation is super-diffusive but with universal scaling dynamics which obey the standard diffusion equation in nonlinearly scaled time. Our results thus reveal a surprising property of an important integrable model as well as pose a challenge to theories which at present are unable to account for our observations. Because the parity symmetry is ubiquitous, our setup should be widely applicable, for instance, we predict a similar physics in the one-dimensional Hubbard model.
	
	\section*{Results}
	
	{\bf The setup.--}
	The Hamiltonian of the $XXZ$ chain of $n$ sites reads
	\begin{equation}
	\label{eq1}
	\mathcal{H}=J\sum_{x=-n/2}^{n/2-1}\left(s_x^{(1)}s_{x+1}^{(1)}+s_x^{(2)}s_{x+1}^{(2)}+\Delta s_x^{(3)}s_{x+1}^{(3)}\right)~, 
	\end{equation}
	where $\Delta$ is the anisotropy parameter and $s_k^{(\gamma)}=\frac{1}{2}\sigma_k^{(\gamma)}$ are the spin $1/2$ operators, with Cartesian component $\gamma=1,2,3$,  expressed in term of Pauli matrices $\sigma_k^{(\gamma)}$ (we use units $J=\hbar=1$). The Hamiltonian preserves the total magnetisation, $M=\sum_x s_x^{(3)}$, $[\mathcal{H},M]=0$.
	We are going to study the spin transport satisfying the continuity equation $d s^{(3)}_x/dt = j_{x-1}-j_x\approx -\nabla j_x$ with the current  
	\begin{equation}
	\label{eq2}
	j_x=s_x^{(1)}s_{x+1}^{(2)}-s_x^{(2)}s_{x+1}^{(1)}~.
	\end{equation}
	\figI
	The existence of spin-reversal parity $S=\prod_x \sigma^{(1)}_x$, $[{\cal H},S]=0$, and odd current $j_x S = -S j_x$, implies an absence of ballistic transport channels based on local conserved charges 
	\cite{zotos}. We are going to simulate the time evolution of an initial inhomogeneous state composed of two halves with opposite magnetisations. 
	\figII
	To this end we choose a product initial state described by a density operator $\rho$, 
	\begin{equation}
	\label{eq3}
	\rho(t=0) \sim \left(1+\mu\, \sigma^{(3)}\right)^{\otimes\frac{n}{2}}\otimes\left(1-\mu\, \sigma^{(3)}\right)^{\otimes\frac{n}{2}}~,
	\end{equation}
	where the parameter $\mu \in [-1,1]$ determines the initial magnetisation, being $\langle s_{x\ge 0,<0}^{(3)}\rangle=\pm\frac{1}{2}\mu$. Each of the initial halves can be thought of as being in equilibrium state $\sim {\rm e}^{\pm h \sum_x s_x^{(3)} }$ at very high temperature and finite magnetisation. We are therefore studying high-energy nonequilibrium physics of the model. While the initial state is pure for $|\mu|=1$ (a fully polarised domain wall), evolution of which has been studied in the past~\cite{Gobert}, the choice of a mixed state offers several important advantages: it is generic and not plagued by the speciality of $\mu=1$ at $\Delta>1$ for which the dynamics freezes due to the proximity to a gapped eigenstate~\cite{Gochev}, and it is, for small $\mu$, better suited for numerical simulations. This allows us to study significantly longer timescales as compared to existing literature and infer the scaling functions. We also mention that such an initial state can be thought of as representing an ensemble of pure states with randomised angle $\varphi$ on the Bloch sphere (see Methods).
	
	{\bf Scaling exponents.--}
	We focus our efforts on $\Delta \ge 1$ where there are no analytic results known for the magnetisation transport, and the method \cite{doyon,bertini} only predicts vanishing ballistic contribution. Two representative examples of a time evolved state $\rho(t)$, namely the spin and current profiles $s(x,t) = {\rm tr} \rho(t) s^{(3)}_x$, $j(x,t) = {\rm tr}\rho(t) j_x$, are shown in Fig.~\ref{fig1}.  
	\figIII
	In order to obtain the exact type of transport we shall quantitatively study equilibration of magnetisation, in particular the scaling of spin and current profiles as well as the transferred magnetisation between the two halves, whose asymptotic scaling power $\alpha$ characterizes the transport type,
	\begin{equation}
	\label{eq4}
	\Delta s(t)=\int_0^tj(0,t'){\rm d}t'\propto t^\alpha~,
	\end{equation}
	where $j(0,t)$ is the current at the half-cut. For $\alpha=1/2$ the transport is diffusive, for $1/2<\alpha<1$ it is called superdiffusive, and finally, $\alpha=1$ corresponds to ballistic transport. We note that the transport type is connected to current-current correlation function via Green-Kubo linear response theory. In case of diffusive transport, the spin density satisfies the diffusion equation. This notion of diffusion does not necessarily correspond to De Gennes phenomenological theory of spin diffusion which, under much stronger assumptions, in one-dimension implies $1/\sqrt{t}$ dependence of local spin density autocorrelation function \cite{affleck11,mccoy}.
	
	We evolved the initial state $\rho(0)$ (\ref{eq3}) up to long times (of order $t\approx 160$) and set large enough $n$ so that there was no significant finite size effects. From the data we then infer the exponent $\alpha$ using Eq.(\ref{eq4}), see Fig.\ref{fig2}(a) and (b) for representative plots. Dependence of the exponent $\alpha$ on $\Delta$ is summarised in Fig.~\ref{fig2}(c). While the transport is found to be ballistic for $\Delta<1$, expectedly so for the integrable system, also known rigorously~\cite{prosen}, at $\Delta\ge 1$ we find rather clear non-ballistic relaxation. In particular, at $\Delta=1$ it is superdiffusive while for $\Delta>1$ the transport is diffusive, observed in driven steady-state setting~\cite{jstat09,prl11} as well as in the Hamiltonian one~\cite{Robin09,FHM,affleck11,robin14,robin17}. At $\Delta=1$ we also observe small dependence of $\alpha$ on $\mu$. While for small $\mu$, i.e., small deviations from an infinite temperature state $\rho\sim\mathbbm{1}$, the exponent is close to $2/3$, closer to pure state $\mu=1$ it appears to be closer to $\approx 3/5$ (we note that a different numerical procedure is used in the two regimes, see Methods). 
	
	{\bf Scaling functions.--}
	The scaling of the transferred magnetisation unequivocally shows a surprising non-ballistic transport in an integrable system which, however, has been observed and discussed before in related contexts, namely within local quench and linear response theory \cite{Robin09,FHM,affleck11,robin14,robin17} and boundary driven Lindblad approach \cite{jstat09,prl11}. But here we can do more still. In Fig.~\ref{fig3} we demonstrate that the spin profiles can be described by a function of a single scaling variable $x/t^\alpha$ -- profiles at large times collapse to a single curve. In addition, the profiles of current and magnetisation are proportional to each other at different times (Fig.~\ref{fig3}(c,d)), therefore validating Fick's law $j = -D\nabla s$ where the behaviour of the diffusion constant $D$ with respect to the anisotropy $\Delta$ is shown in the inset of Fig.~\ref{fig2}(c). This comes as no surprise in the diffusive regime $\Delta>1$ where the scaling function of the magnetisation (Fig.~\ref{fig3}(b)) is simply the error function $s(x,t) = -\frac{\mu}{2}\text{erf}(x/\sqrt{4Dt})$. However, the same can not be said for the isotropic point $\Delta=1$. Proportionality between the magnetisation gradient and the current profile (Fig.\ref{fig3}(c)), this time with a time-dependent ratio $D\simeq \frac{K}{3} t^{1/3}$, suggests a diffusion equation in a scaled time
	\begin{equation}
	\label{eq5}
	\frac{\partial s(x,t)}{\partial \tau}= \frac{K}{4} \frac{\partial^2 s(x,t)}{\partial x^2}~,\quad \text{where}\quad \tau = t^{4/3},
	\end{equation}
	which again yields error function profile with a different scaling variable $s(x,t) = -\frac{\mu}{2}\text{erf}(K^{-1/2}x/t^{2/3})$ with $K=2.33\pm0.03$.
	In Fig.\ref{fig3}(a) we compare numerical profiles with the error function, again finding good agreement within accuracy of our simulations. Therefore, the scaling function is, in both cases, $\Delta=1$ and $\Delta>1$, the error function, the difference being only in the scaling variable which is $x/t^{2/3}$ at the superdiffusive isotropic point. This result is surprising, as anomalous diffusion is usually associated with Levy processes and hence long (non-Gaussian) tails in the profiles. Here it seems it all amounts to a nonlinear rescaling of time. Theoretical explanation of this effect is urgent.
	
	\figIV
	
	{\bf Entanglement entropy and simulation complexity.--}
	Lastly, we mention a numerical observation that explains why we can simulate dynamics to such long times, and is an interesting property on its own.
	We use a time-dependent density matrix renormalisation group method (tDMRG), see Methods.
	The efficiency of tDMRG depends on the entanglement entropy, i.e., for pure state evolution on the Von Neumann entropy $S=-{\rm tr}[\rho_A \ln{\rho_A}]$ of the reduced state $\rho_A = \tr_A |\Psi\rangle \langle\Psi|$, whereas for mixed states evolution on an analogous operator space entanglement entropy $S^\#$ \cite{ProsenPizorn} of a vectorised density operator $\rho$.	When starting with a typical product initial state both entropies typically grow linearly with time, regardless of the system being integrable or not~\cite{CalabreseCardy,Chiara06}, causing exponentially fast growth of complexity and with it a failure of these numerical methods. In our case though, see Fig.\ref{fig4}, entropies grow much slower, namely in a power-law fashion 
	\begin{equation}
	\label{eq6}
	S\sim t^\beta, \quad \text{or}\quad S^{\#} \sim t^\beta,
	\end{equation}
	with $\beta$ being less than $1$. The most efficient simulations have been possible with density operators for small $\mu$ where the exponent $\beta$ is typically between $0.3$ and $0.5$.
	
	\section{Discussion}
	
	Our numerical results can be interpreted as an evidence of normal spin-diffusion and spin Fick's law in the easy-axis anisotropic Heisenberg chain (for anisotropy $\Delta > 1$), with spin-density satisfying the diffusion equation on large scales. Besides the case $\Delta=2$ shown here, we provide additional data for $\Delta=1.05, 1.1, 1.3, 1.5$ demonstrating a clear convergence of the diffusive scaling exponents $\alpha=1/2$ in all massive cases (Supplementary Note 1), and data for massless cases $\Delta=0,0.5,0.7,0.9$ which indicate convergence to ballistic exponent $\alpha=1$ (Supplementary Note 2). While for generic, non-spin-reversal-symmetric initial states, the dominant contribution to transport is ballistic as determined by generalised hydrodynamics (or generalised one-dimensional Euler's equations) \cite{doyon,bertini,enej,moore,doyonspohn}, the next-to-leading term is now clearly predicted to be diffusive, as following from our work. However, a theoretical explanation, or even derivation of diffusive contribution to transport in an integrable system with a macroscopic number of conservation laws is still pending. Even more surprising is the discovery of anomalous super-ballistic transport in the isotropic case ($\Delta=1$) with the scaling exponent equal to or very close to $2/3$. While this might suggest a behaviour described by KPZ (Kardar-Parisi-Zhang) universality class, we find that asymptotic spin density profiles obey the non-linearly scaled diffusion equation and are distinct from the KPZ profiles. One might conjecture that the scaling exponent $2/3$ is a consequence of $SU(2)$ symmetry and not the fact that the model there corresponds to the marginal critical point $\Delta=1$. This would be consistent with observed anomalous super-diffusive scalings in $SU(4)$ spin 
	ladders in the setup of driven steady state Lindblad dynamics \cite{spinladder} where the scaling exponent appears to be $\alpha=3/5$. Curiously, all scaling exponents observed in this work ($1/1$, $1/2$, $2/3$, $3/5$) are ratios of subsequent Fibonacci numbers \cite{popkov}.
	
	\section{Methods}
	
	{\bf Numerical procedures.--} The time evolution is performed by means of the tDMRG algorithm~\cite{vidal, schollwock}.
	In particular, for small $\mu$ data (which is mostly reported here) the most efficient was the matrix product density operator version of tDMRG, with which we could reach times of the order $t\simeq 200$ for system size $n\simeq 2t$ using bond dimensions $50-200$ resulting in relative truncation errors less than $1\%$.
	One the other hand, for $\mu\approx 1$ (close to domain wall pure state), the pure state version of tDMRG becomes more efficient as the corresponding entanglement entropy scaling exponents $\beta$ are smaller.
	The two approaches appear to complement one another as can be seen in Fig.\ref{fig2}(d). 
	Neither approach allows us to observe long times in the intermediate region of $\mu$, where the exponents $\beta$ become closer to $1$.
	
	In order to simulate the desired density operator by evolving pure states we define a set of initial states 
	\begin{equation}
	\label{eq7}
	|\Psi(t=0)\rangle=\bigotimes_{x < 0}|\psi(\mu,\phi_x)\rangle \otimes \bigotimes_{x \ge 0}|\psi(-\mu,\phi_x)\rangle
	\end{equation}
	where $|\psi(\mu,\phi)\rangle=\sqrt{(1+\mu)/2}|\hspace{-1mm}\uparrow\rangle+e^{i\phi}\sqrt{(1-\mu)/2}|\hspace{-1mm}\downarrow\rangle$ is simply the Bloch sphere representation of a 2-level system and the $\phi_x$ are uniform independent random numbers in the range $[0,2\pi)$. The density matrix is then obtained as an ensemble average over a set of such pure random states $\rho(t) = \mathbb{E}(|\Psi(t)\rangle \langle\Psi(t)|)$. It is clear that an increasingly large set of random states is needed as the magnetisation approaches $\mu\to0$, where the matrix product density operator simulation is  favourable anyway. 
	
	\section{Acknowledgements}
	
	The authors would like to acknowledge support from ERC grant OMNES, as well as grants  J1-7279, N1-0025, and P1-0044 of the Slovenian Research Agency.

	\appendices
	
	\section{Supplementary Note 1 - Diffusive Regime}

\begin{figure}[ht!]
		\centering
		\includegraphics[width=8.5cm]{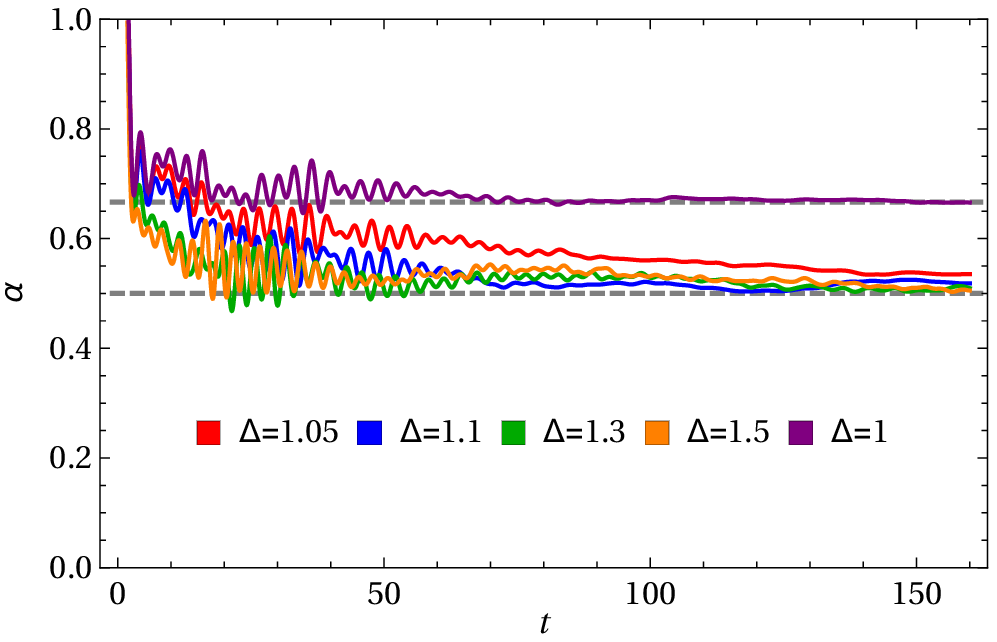}
		\caption{\textbf{Scaling exponents. } We show the time dependence of the scaling exponent $\alpha$ for various values of the anisotropy $\Delta$ in the massive regime, as well as for the isotropic reigme $\Delta=1$ for comparison. 
			Nearer to $\Delta=1$ the time scale at which the final value is reached appears to increase but the convergence trend is clear. }
		\label{sfig1}
	\end{figure}

	We direct the readers attention to Supplementary Figure \ref{sfig1}, where the local-time exponent $\alpha$ is shown as a function of time for several values of the anisotropy in the massive regime, namely $\Delta=1.05,1.1,1.3,1.5$. The asymptotic scaling exponent converges on the accessible time scale to $\alpha=1/2$ -- expect for the case $\Delta=1.05$ where the convergence has not yet been reached though the extrapolated asymptotic value is likely the same -- is distinctly different from the exponent $\alpha=2/3$ for the isotropic case $\Delta=1$.
	
	\section{Supplementary Note 2 - Ballistic Regime}	
	
	Spin transport is known to be ballistic in the massless regime, $\Delta<1$. Here we verify that the chosen inhomogeneous quench does indeed reproduce this result.  
	In Supplementary Figure \ref{sfig2} we demonstrate the convergence of local-time scaling exponent $\alpha$ for a few cases $\Delta=0,0.5, 0.7,0.9$ to an asymptotic ballistic exponent $\alpha=1$, and again compare it to the isotropic case $\Delta=1$.
	
	\begin{figure}[ht!]
		\centering
		\includegraphics[width=8.5cm]{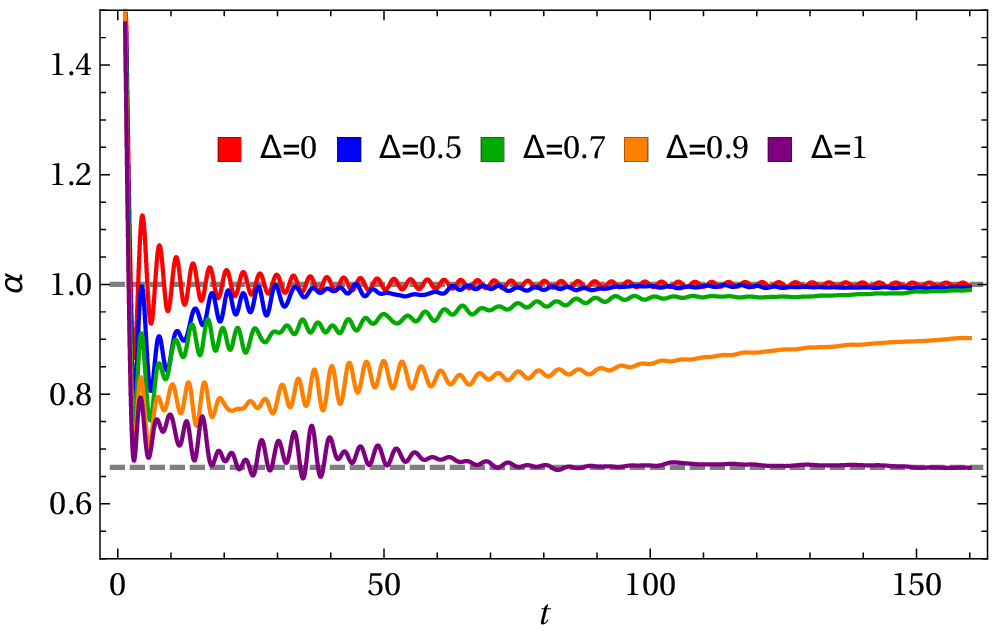}
		\caption{\textbf{Scaling exponents. } We show the time dependence of the scaling exponent $\alpha$ for various values of the anisotropy $\Delta$ in the massless regime. We observe clear asymptotic convergence towards the ballistic value $\alpha$, while the convergence times increase when $\Delta$ approaches $1$, however the transition to $\Delta=1$ -- shown for comparision -- appears to be discontinuous. }
		\label{sfig2}
	\end{figure}
	
	In Supplementary Figure \ref{sfig3} we also show the scaling of spin and current density profiles which clearly exhibit expected ballistic behaviour \cite{doyon,bertini}.  In the non-interacting case $\Delta=0$ we find also excellent agreement with analytic solutions \cite{antal}.  

\widetext

	\begin{figure}[ht!]
		\centering
		\includegraphics[width=0.49\textwidth]{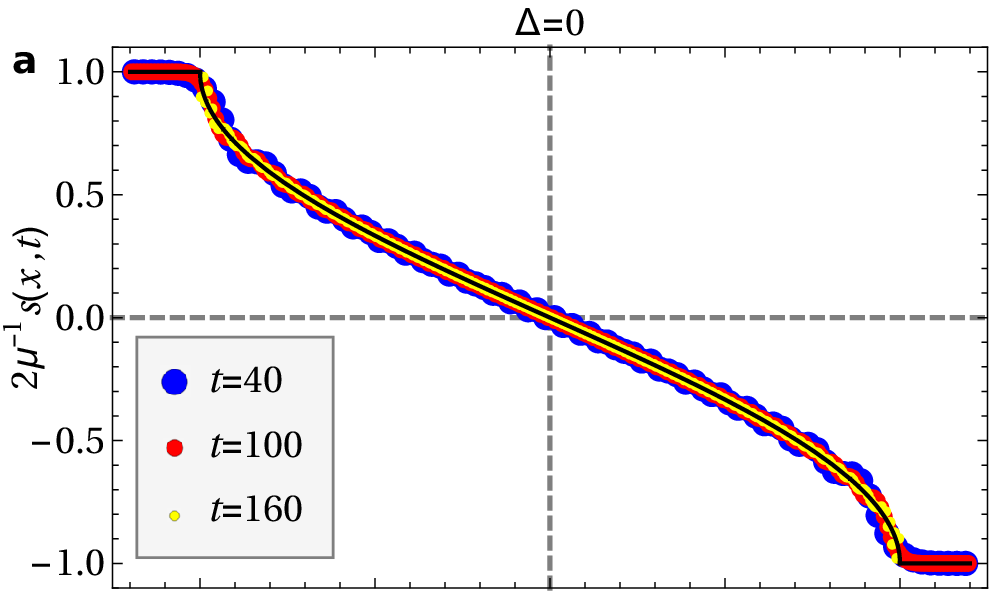}
		\includegraphics[width=0.49\textwidth]{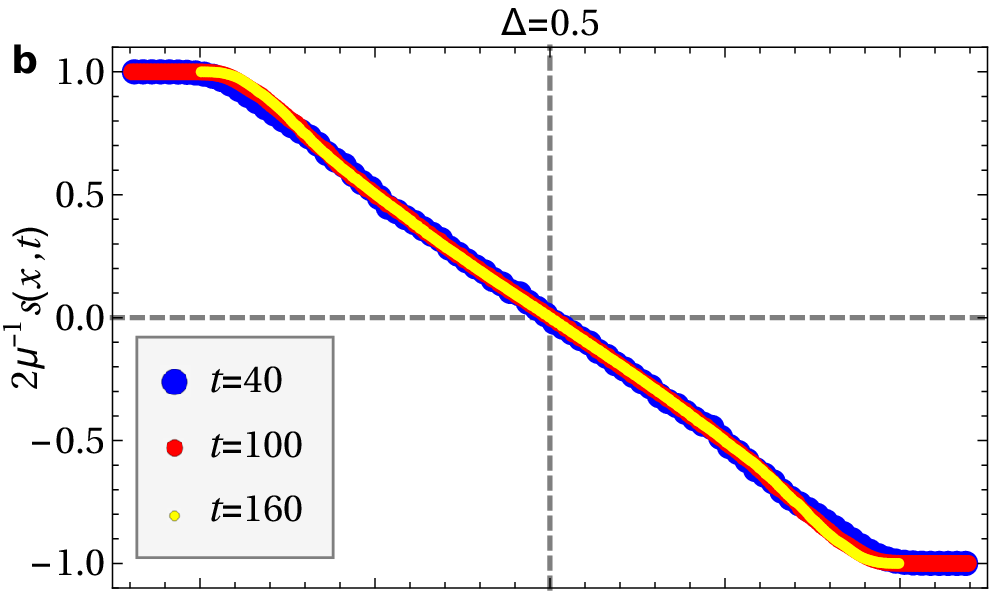}
		\includegraphics[width=0.49\textwidth]{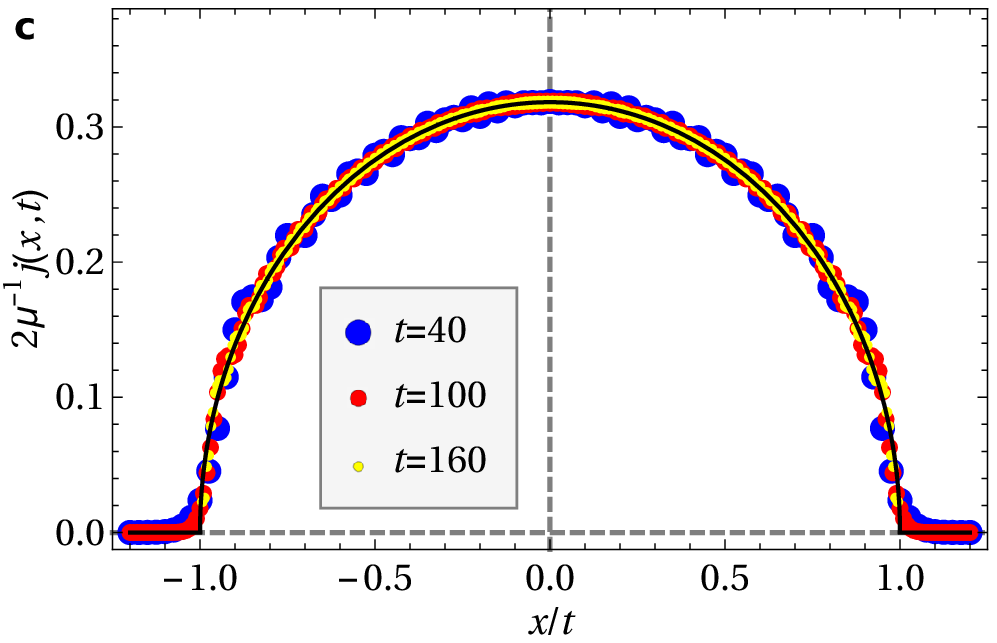}
		\includegraphics[width=0.49\textwidth]{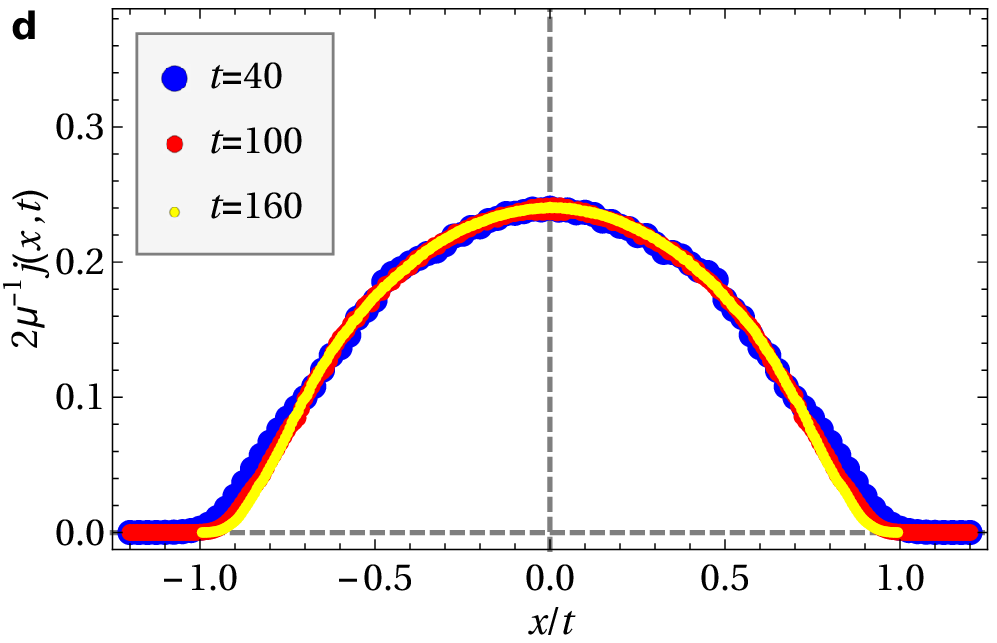}
		\caption{\textbf{Scaling functions. } We show spin (a and b) and current (c and d) density profiles for two values of the anisotropy $\Delta=0$ (left) and $\Delta=0.5$ (right) with respect to a single scaling variable $x/t$ suggestive of ballistic transport expected for this regime. In the non-interacting case $\Delta=0$ we also indicate
			known analytic solutions which excellently match the numerical data.}
		\label{sfig3}
	\end{figure}

\end{document}